# Stress evolution in plastically deformed austenitic and ferritic steels determined using angle- and energy-dispersive diffraction

M. Marciszko-Wiąckowska[1,*], A. Baczmański[2], Ch. Braham[3], M. Wątroba[4], S. Wroński[2], R. Wawszczak[2], G. Gonzalez[5], P. Kot[6], M. Klaus[7], Ch. Genzel[7]

[1]AGH University of Krakow, ACMIN, al. Mickiewicza 30, 30-059 Kraków, Poland

[2]AGH University of Krakow, WFiIS, al. Mickiewicza 30, 30-059 Kraków, Poland

[3]Arts et Métiers-ParisTech, PIMM, CNRS UMR 8006, 151 Bd de l'Hôpital, 75013 Paris, France

[4]EMPA, Swiss Federal Laboratories for Materials Science and Technology, Labratory for Mechanics of Materials and Nanostructures, Feuerwerkerstrasse 39, 3602 Thun, Switzerland

[5]Instituto de Investigaciones en Materiales, Universidad Nacional Autónoma de México, Circuito Exterior S/N, Cd. Universitaria, A.P. 70-360, Coyoacán, C.P. 04510, Mexico

[6]NOMATEN Centre of Excellence, National Centre of Nuclear Research, A. Sołtana 7, 05-400 Otwock-Świerk, Poland

[7]Abteilung fürMikrostruktur- und Eigenspannungsanalyse, Helmholtz-Zentrum Berlin fürMaterialien und Energie, Albert-Einstein-Str. 15, Berlin 12489, Germany

*Corresponding author: Marianna Marciszko-Wiąckowska (marciszk@agh.edu.pl), +48126175309

**Abstract**

In the presented research, the intergranular elastic interaction and the second-order plastic incompatibility stress in textured ferritic and austenitic steels were investigated by means of diffraction. The lattice strains were measured inside the samples by the multiple reflection method using high energy X-rays diffraction during uniaxial in situ tensile tests. Comparing experiment with various models of intergranular interaction, it was found that the Eshelby-

Kröner model correctly approximates the X-ray stress factors (XSFs) for different reflections hkl and scattering vector orientations.

The verified XSFs were used to investigate the evolution of the first and second-order stresses in both austenitic and ferritic steels. It was shown that considering only the elastic anisotropy, the non-linearity of $\sin^2\psi$ plots cannot be explained by crystallographic texture. Therefore, a more advanced method based on elastic-plastic self-consistent modeling (EPSC) is required for the analysis. Using such methodology the non-linearities of $\cos^2\varphi$ plots were explained, and the evolutions of the first and second-order stresses were determined. It was found that plastic deformation of about 1- 2% can completely exchange the state of second-order plastic incompatibility stresses.

## 1. Introduction

Knowledge of the residual stress state is essential to understanding the mechanical behavior of polycrystalline materials. The element may be damaged or strengthened when external stresses are added to this stress, which develops during mechanical or thermal processing. Therefore, the macroscopic residual stress (first-order stress) in the element's subsurface layer is important; for instance, compressive stress slows crack initiation and propagation, while tensile stress typically speeds it up. Also, the so-called second-order residual stresses [1], characterizing the heterogeneity of the stresses on the scale of polycrystalline grains, may affect the plastic deformation process of the material [2,3]. These stresses, defined as the deviations of the stresses for individual polycrystalline grains from the mean macroscopic value (first-order stress), are caused by anisotropy or heterogeneity of the processes occurring for different grains and, as a result, lead to a mismatch in their shape or volume. For example, during plastic deformation, second-order stresses result from the difference in activation of the slip system or twin phenomena occurring in grains belonging to different phases or having different lattice orientations [2–9].

In recent years, much effort has been put into studying the phenomena occurring in individual grains using synchrotron diffraction measurements (e.g. [10–17]). These measurements make it possible to determine the state of stress for individual grains during the deformation of the sample. For example, in the work [10], the stress state in the grains of a titanium sample was determined using high-energy X-ray diffraction microscopy. However, the tests were carried out for a limited number of single grains, and some dispersion of results was found. The first direct determination of the stress state for different grains in the AZ31 magnesium alloy was carried out by neutron diffraction [18]. In situ, measurements were made for several preferred texture orientations. Such experiments are promising due to the high statistics of the grains for which the measurement is performed. However, both methods have some limitations; direct grain stress scanning with a microscopic synchrotron beam can be performed for large grains with dimensions of several tens of μm, and due to the long measurement time and complicated data processing, it can be performed for a limited number of specific grains. On the other hand, neutron measurements of grain stresses for groups of grains can only be performed for a highly textured sample.

Diffraction methods based on the measurement of lattice deformations are commonly used to determine first-order stresses in a polycrystalline material, both textured and quasi-isotropic (i.e. having a random orientation distribution). The stress tensor can be calculated from the measured elastic strains only when the so-called XECs or XSFs (X-ray Elastic Constants or X-ray Stress Factors [19,20]) relating lattice elastic deformation with the first-order stresses (mean stresses for considered volume), are known. The values of XSFs (or XECs) are determined directly from the experiment or calculated from well-known grain interaction models [9,19,21], e.g., Reuss [22], Voigt [23], Eshelby-Kröner [24–26], and free-surface [9]. The concepts of the free-surface model are provided in [9], whereas the standard models, i.e., Reuss, Voigt, and Eshelby- Kröner approaches, are extensively discussed in the literature (e.g. [19–21,27]). However, a given model's applicability is usually not well-argued, considering such effects as relaxation of the forces perpendicular to the surface, grains size, and shape. In the previous works, usually, the Eshelby-Kröner was regarded as the closes to the real material; however, it is not a general rule as was shown, for example, in the case of near-surface volume [28,29], in the case of textured samples [27,30] or for the columnar microstructure of grains in the coatings [21]. Therefore, the first objective of this study is to verify the accuracy of XSF's model in stress analysis using representative information gained from the experiment, i.e., using different *hkl* reflections and many orientations of the scattering vector. So, the Voigt, Reuss, and Eshelby-Kröner models are compared with experimental XSFs determined for austenitic and ferritic steel. A further requirement for the model's applicability is the accuracy of the fit between the theoretical and measured lattice strains.

The important aim of the present study is to apply the verified XSFs to study the development of the first- and the second-order stress in textured austenitic and ferritic steel subjected to a uniaxial tensile test. The stress was determined using the multireflection method [3,31–33]. Second-order plastic incompatibility stresses are the fluctuation around mean stress resulting from differences in plastic deformation of crystals having different orientations. Their effect may be observed as non-linearities or changes in the slope of the $<a(\psi,\varphi)>_{hkl}$ vs. $sin^2\psi$ for different orientations of the scattering vector (where $\psi$ angle is between scattering vector and normal to the surface and $\varphi$ is a rotation angle about the normal, c.f. [30] measured for polycrystalline samples subjected to elastoplastic deformation [34–36]. The elastic anisotropy of the crystallites within textured samples would be the first cause of the non-linearities; however, in many cases, the effect of the second-order stresses can be even more significant [8,9,31]. The elastic anisotropy can be incorporated in stress analysis if the XSF values are known from the experiment or computed using the suitable grain interaction model, as has been done in this work. The calculations of XSFs, based on the crystallographic texture and single crystal elastic constants, allow for the prediction of the character of $sin^2\psi$ plots for a sample under applied or residual macrostresses. Nonetheless, the interpretation of non-linearities based solely on elastic anisotropy is typically insufficient. Therefore, the first- and second-order plastic incompatibility stresses analysis requires a more advanced method based on modeling the plastic deformation process [9,25,37–41]. It is worth noting that in many works, the presence of the second-order plastic incompatibility stresses was observed as the changes in the tendency of the lattice strains dependence vs. macroscopic stress, measured in the direction of applied load and transverse direction (the experimental lattice strains were compared with elastic-plastic self-consistent model) [40,42]. However, the distribution of these stresses in Euler space or their mean amplitude was not determined. Previously, we determined the mean von Mises second-order plastic incompatibility stresses using multiple reflection methods during tensile for duplex steel [3] and pearlitic steel [32]. A similar analysis is done in this study, but additionally the distribution of second-order stresses in Euler space for single-phase austenitic and ferritic steels is determined and presented. The methodology presented below was applied to in situ measurements of lattice strains during a tensile test. The diffraction experiment was performed in transmission mode using high-energy synchrotron radiation for austenitic and ferritic steel. The XSFs were calculated using the Eshelby-Kröner model, accounting for texture [21], and the model values were verified in this experiment during the unloading of the sample. Hence, the complete stress state analysis has been done using correctly determined XSFs and including the second-order stresses.

The basic equation relating the first-order stresses $\sigma_{ij}^I$ with corresponding elastic lattice strain $<\varepsilon(\psi,\varphi)>_{hkl}^e$ can be written in the following form:

$$\frac{<d(\psi,\varphi)>_{hkl}^\sigma - d_{hkl}^0}{d_{hkl}^0} = <\varepsilon(\psi,\varphi)>_{hkl}^e = F_{ij}(hkl,\psi,\varphi,f)\sigma_{ij}^I \quad (1)$$

where: $<d(\psi,\varphi)>_{hkl}^\sigma$ is the values of interplanar spacings measured using *hkl* reflection in a direction characterized by angles ψ and φ [19,30] for a material subjected to the first-order stress $\sigma_{ij}^I$, $d_{hkl}^0$ is the stress free interplanar spacing, $F_{ij}(hkl,\psi,\varphi,f)$ are the XSFs and $f$ denotes ODF (Orientation Distribution Function) characterizing crystallographic texture.

This equation represents an elastic response of the lattice to the residual or applied stress $\sigma_{ij}^I$ and it can be used to determine the values of XSFs experimentally. For example, if a known increment of uniaxial stress $\Delta\Sigma_{11} = \Sigma_{11}^{(2)} - \Sigma_{11}^{(1)}$ applied to the sample during the tensile test, the factor $F_{11}(hkl,\psi,\varphi,f)$ can be calculated from the corresponding lattice strain, measured as the change in the interplanar spacings:

$$F_{11}(hkl,\psi,\varphi,f) = \frac{<\varepsilon(\psi,\varphi)>_{hkl}^e}{\Delta\Sigma_{11}} = \frac{<d(\psi,\varphi)>_{hkl}^{\Sigma 2} - <d(\psi,\varphi)>_{hkl}^{\Sigma 1}}{<d(\psi,\varphi)>_{hkl}^{\Sigma 1}\Delta\Sigma_{11}} \quad (2)$$

where: $<d(\psi,\varphi)>_{hkl}^{\Sigma 2}$ and $<d(\psi,\varphi)>_{hkl}^{\Sigma 1}$ correspond respectively to the interplanar spacings measured under uniaxial loads $\Sigma_{11}^{(2)}$ and $\Sigma_{11}^{(1)}$ applied to the sample.

The above formula enables the determination of the XFSs from measured in situ changes in interplanar spacings using appropriate experimental techniques for the increments of the stresses $\Delta\Sigma_{11}$ in the elastic range of sample loading or unloading. Then, the model calculated XSFs $F_{11}^{mod}(hkl,\psi,\varphi,f)$ can be compared with the experimental ones $F_{11}^{exp}(hkl,\psi,\varphi,f)$. It should be emphasized that in this incremental method, the influence of the residual stresses on the experimentally determined XSFs is avoided because the lattice strains corresponding to these stresses are canceled when the differences $< d(\psi, \varphi) >_{hkl}^{\Sigma2} - < d(\psi, \varphi) >_{hkl}^{\Sigma1}$ are calculated [28,43].

Verification of the XSFs, presented in our previous work [28], can be done by analyzing the results of residual stress measurements. In such a case, the accuracy of fitting theoretical lattice parameter values to measured ones is evaluated for several XSFs models. Assuming that the effect of the $\sigma_{ij}^I$ on the measured $< a(\psi, \varphi) >_{hkl}^{\sigma}$ is dominating (i.e., the plastic incompatibility is not significant, and the influence of stacking faults is negligible), the values of the first-order stresses $\sigma_{ij}^I$ and value of strain-free parameter $a_0$ can be determined using least square method based on the well-known equation:

$$< a(\psi, \varphi) >_{hkl}^{\sigma} = F_{ij}(hkl, \psi, \varphi, f)\sigma_{ij}^I a_0 + a_0 \tag{3}$$

where the values $< a(\psi, \varphi) >_{hkl}^{\sigma} = \sqrt{h^2 + k^2 + l^2} < d(\psi, \varphi) >_{hkl}^{\sigma}$ are determined experimentally for cubic lattice.

The verified XSFs can be used to determine residual or imposed sample stresses. In this work, the Eshelby-Kröner XSF model is applied to study the variation of the stress state within the austenitic and ferritic sample during elastic-plastic deformation. The diffraction measurements of the lattice strains were done inside the sample (transmission method) during a tensile test.

In a plastically deformed material, the lattice strains $< \varepsilon(\psi, \varphi) >_{hkl}$ can be expressed as a superposition of strains induced by macrostresses $< \varepsilon(\psi, \varphi) >_{hkl}^{e}$ and the lattice strains induced by second-order incompatibility stresses $< \varepsilon(\psi, \varphi) >_{hkl}^{pi}$:

$$< \varepsilon(\psi, \varphi) >_{hkl} = < \varepsilon(\psi, \varphi) >_{hkl}^{e} + < \varepsilon(\psi, \varphi) >_{hkl}^{pi} = \frac{<a(\psi,\varphi)>_{hkl} - a_0}{a_0} \tag{4}$$

where: $< \varepsilon(\psi, \varphi) >_{hkl}^{e} = F_{ij}(hkl, \psi, \varphi)\sigma_{ij}^I$.

The elastic-plastic self-consistent (EPSC) model has already been presented and proven to determine both types of stresses [41,44,45]. In this method, it is assumed that $< \varepsilon(\psi, \varphi) >_{hkl}^{pi} =$

$q < \tilde{\varepsilon}(\psi,\varphi) >_{hkl}^{pi}$, where $< \tilde{\varepsilon}(\psi,\varphi) >_{hkl}^{pi}$ can be calculated by the self-consistent model and $q$ is a fitting parameter scaling the magnitude of plastic strains.

Therefore, the experimental lattice parameters $< a(\psi,\varphi) >_{hkl}$ obtained from the diffraction method can be expressed as (cf. Eq. 4) [3,5,8,9,46,47]:

$$< a(\psi,\varphi) >_{hkl} = \left[F_{ij}(hkl,\psi,\varphi,f)\sigma_{ij}^I + q < \tilde{\varepsilon}(\psi,\varphi) >_{hkl}^{pi}\right]a_0 + a_0 \quad (5)$$

In the present interpretation, both terms of Eq. 4 are considered. Next, using information from the EPSC model, the magnitude of the residual stress $\sigma_{ij}^{II,pi}$ and its dependence on the crystal orientation may be determined:

$$\sigma_{ij}^{II,pi} = q\, \tilde{\sigma}_{ij}^{II,pi} \quad (6)$$

where $\tilde{\sigma}_{ij}^{II,pi}$ are the plastic incompatibility stresses calculated from the EPSC model.

In the case when diffraction elastic constants are known, strains are theoretically predicted, and lattice spacings are measured, all other unknown quantities from Eq. 5 can be found using the least square procedure, based on minimizing the merit function called $\chi^2$, which is defined as:

$$\chi^2 = \frac{1}{N-M}\sum_{n=1}^{N}\left(\frac{<a(\psi,\varphi)>_{hkl}^{\sigma,exp} - <a(\psi,\varphi)>_{hkl}^{\sigma,cal}}{\delta_n}\right)^2 \quad (7)$$

where $< a(\psi,\varphi) >_{hkl}^{\sigma,exp}$ and $< a(\psi,\varphi) >_{hkl}^{\sigma,cal}$ are the experimental and calculated lattice parameters, respectively, $\delta_n = (< a(\psi,\varphi) >_{hkl}^{\sigma,exp})$ is the measurement uncertainty (standard deviation) of $< a(\psi,\varphi) >_{hkl}^{\sigma,exp}$ for the *n-th* measurement, *N* and *M* are the numbers of measured lattice parameters and fitting parameters, respectively.

The described methodology is applied in the presented study to simultaneously calculate first-order stresses and plastic incompatibility second-order stresses in the single-phase austenitic and ferritic steel using energy- and angle-dispersive diffraction. To predict the theoretical second-order stresses caused by plastic incompatibilities, the Elasto-Plastic Self-Consistent (EPSC) model based on the work of Berveiller and Lipiński [44] was used.

In model prediction, the sample is represented by a number of grains, having a distribution of orientations reproducing the initial experimental textures. First, the model sample is subjected to elastoplastic deformation; next, the external stresses are unloaded. So performed modeling was carried out to determine the values of $\tilde{\varepsilon}(\psi, \varphi)$, which are used in stress analysis using Eq. 5.

## 2. Experimental
### 2.1. Material

The presented study investigated two materials, austenitic and ferritic stainless steel (composition given in Table 1), with high elastic anisotropy (Zener ratio A, given in Table 2). The single-crystal elastic constants ($C_{ij}$) used in this work for the investigated samples are gathered in Table 2.

Table 1. Composition of investigated stainless steals samples (wt.%) (ASS: austenitic stainless steal 316L (Z2CND17-12) and FSS: ferritic stainless steal AIPI 5L X65

|  | Fe | Cr | Ni | Mo | Mn | Cu | Si | P | S | C | N | Co |
|---|---|---|---|---|---|---|---|---|---|---|---|---|
| **ASS** | bal. | 16.63 | 11.14 | 2.03 | 1.31 | 0.35 | 0.52 | 0.022 | 0.025 | 0.02 | 0.03 | 0.18 |
| **FSS** | bal. | 0.034 | 0.399 | 0.023 | 1.38 | 0.193 | 0.285 | 0.008 | 0.001 | 0.031 | -- | -- |

Table 2. Single crystal elastic constants (used for XSF calculation) [48] together with the Zener ratio.

| Material | $C_{11}$ (GPa) | $C_{12}$ (GPa) | $C_{44}$ (GPa) | A |
|---|---|---|---|---|
| Fe-austenite | 197 | 122 | 124 | 3.3 |
| Fe-ferrite | 231 | 134.4 | 116.4 | 2.4 |

A dog-bone-shaped tensile specimen with the following gauge dimensions: 5 mm in width, 3 mm in thickness, and 33 mm in length was made of a hot-rolled austenitic steel sheet. In the case of a cold-rolled ferritic steel sheet, the dimensions of the sample having a similar shape were 1.5 mm in width, 1.5 mm in thickness, and 12 mm in length. The initial microstructure was analyzed using a Tescan Mira scanning electron microscope (SEM) equipped with an EDAX DigiView electron back-scattered diffraction (EBSD) camera. An EBSD analysis was performed on two cross-sections of each specimen, perpendicular to the rolling direction and perpendicular to the normal direction (Fig.1). The samples were prepared for microscopic observations using standard metallographic preparation steps. EBSD maps were collected at 25 kV, from 300 μm x 300 μm area with a step size of 0.25 μm. The average grain size, crystallographic orientation maps were analyzed using TSL OIM™ Analysis software. A single grain was defined as a set of at least 5 measurement points surrounded by a continuous grain boundary segment with a misorientation of at least 15˚. In Fig. 1 the Inverse Pole Figure (IPF) maps for austenitic and ferritic samples were presented. A uniform microstructure with approximately equiaxed recrystallized grains in the austenitic sample with some recrystallization twins is seen. In the case of the ferritic sample, the grains show defected microstructure and more complex shapes but without significant elongation in one direction. Therefore in model calculations, the grains in both phases were approximated by spherical Eshelby inclusions, exhibiting statistically isotropic interaction with the matrix. The average grain size in austenitic and ferritic sample is $14.2 \pm 7.5$ μm and $9.3 \pm 5.9$ μm, respectively.

The crystallographic texture was characterized by the X-ray diffraction method using Co radiation (Empyrean XRD Diffractometer, Malvern Panalytical). To do this, the pole figures 110, 200, 211 for ferrite and 111, 200, 220 for austenite were measured. The ODFs calculated from pole figures using the WIMV method [49] are shown in Fig. 2., where the orientation of the sample coordinate system respectively to the main directions of the rolling process is defined.

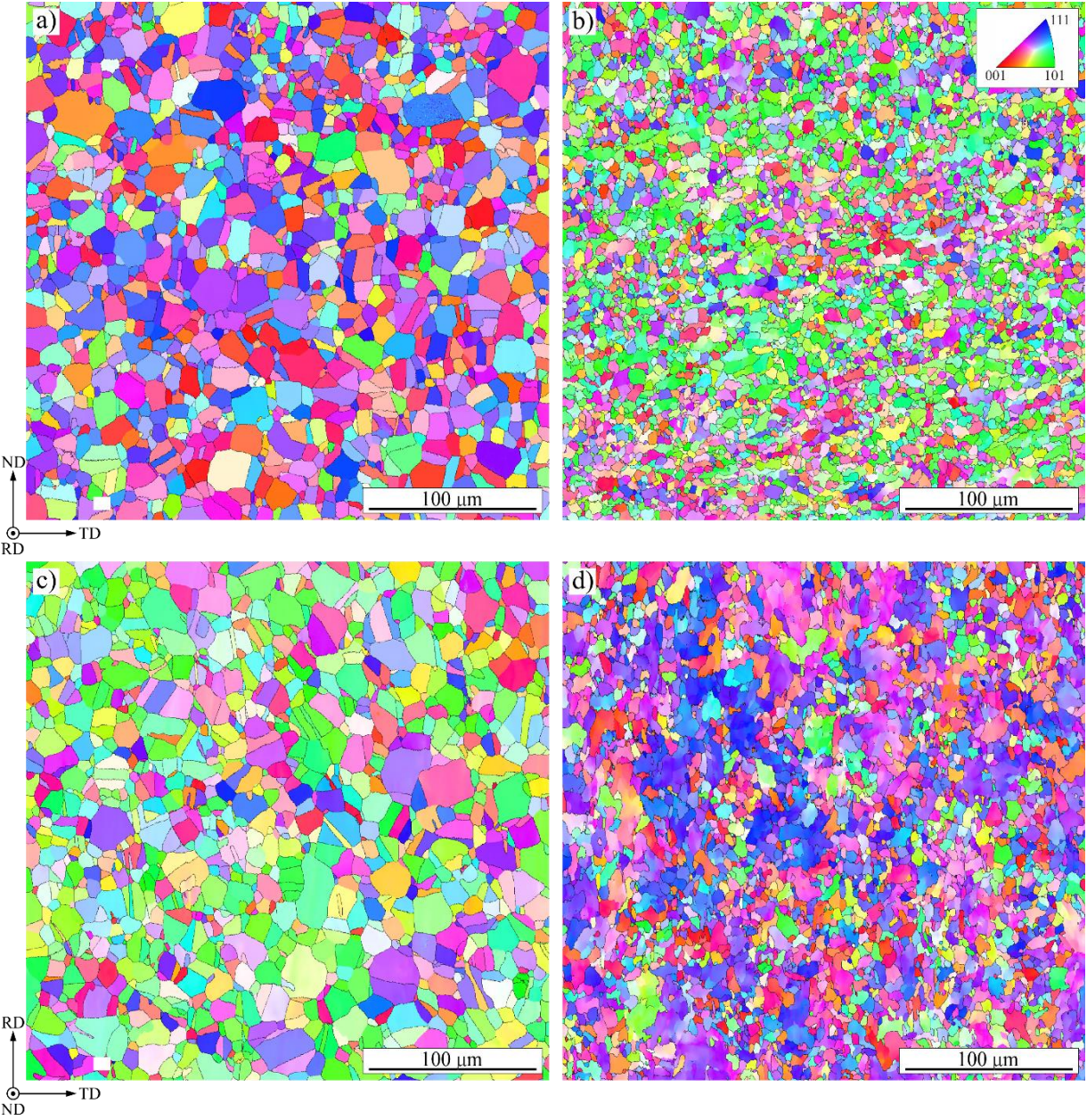

Fig. 1. The EBSD-IPF orientation maps show the microstructure of austenitic (a, c) and ferritic (b, d) steel samples. The maps were measured at two different cross sections: (a, b) plane determined by rolling direction (RD) and normal direction (ND); (c, d) plane determined by rolling direction (RD) and transverse direction (TD).

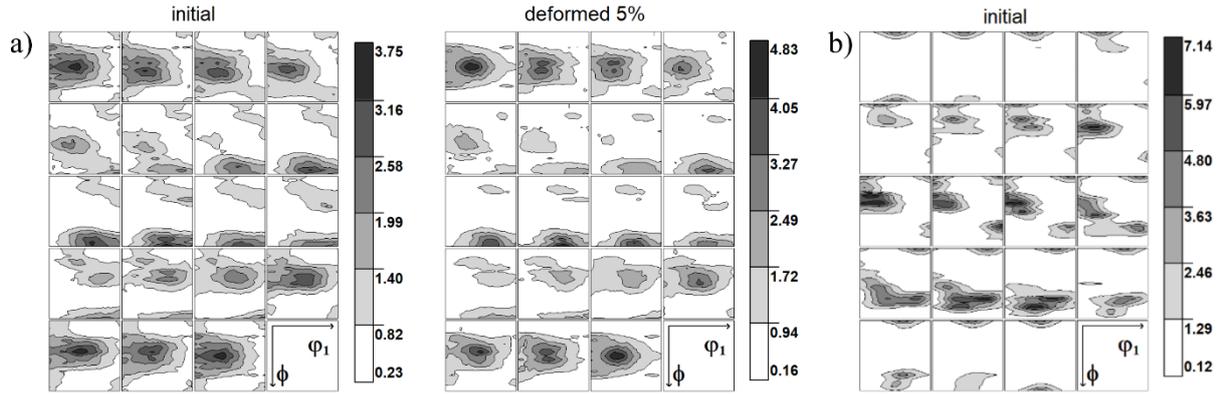

Fig. 2. Normalized orientation distribution function (ODF) determined using Co radiation for austenitic (a) and ferritic (b) steel. For ferritic steel, only the initial texture was presented as the deformation for this sample was small. The sections through Euler space with the step of 5° are presented along the $\phi_2$. The Euler angles are defined with respect to sample axes RD, TD and ND (as in standard presentation for rolled sheet [50]). The levels express multiples-of-random-distribution.

## 2.2. Measurements

The X-ray diffraction method was used to measure in situ lattice strains during the tensile test. Two separate measuring methods: energy-dispersive (ED) diffraction for austenitic steel and angle-dispersive (AD) diffraction for ferritic steel, were used to determine the lattice strains in situ during a tensile test. As presented in Fig. 3, an experimental setup contained a dog-bone-shaped specimen stretched along the transverse direction (TD). Then, during the tensile test, the actual deformation in the elastic range and for small plastic deformation was measured by a gauge placed on the sample.

It should be emphasized that during deformation of austenitic 316L stainless steel, a dynamic phase transitions and twinning process can occur. However, in the diffraction patterns collected during the entire experiment, only diffraction peaks corresponding to the pure austenitic phase were found, which means that even if transformations occurred, the volume fractions of the new phases are insignificant compared to the austenitic phase. In addition, twinning is more likely at higher strains than those used in our experiment, i.e. above 5 - 10% [51–53]. Therefore, it was assumed that crystallographic slip is the dominant deformation process. The same assumption was made for ferritic steel, for which the presence of a pure ferritic phase was found in the diffractograms collected during the tensile test.

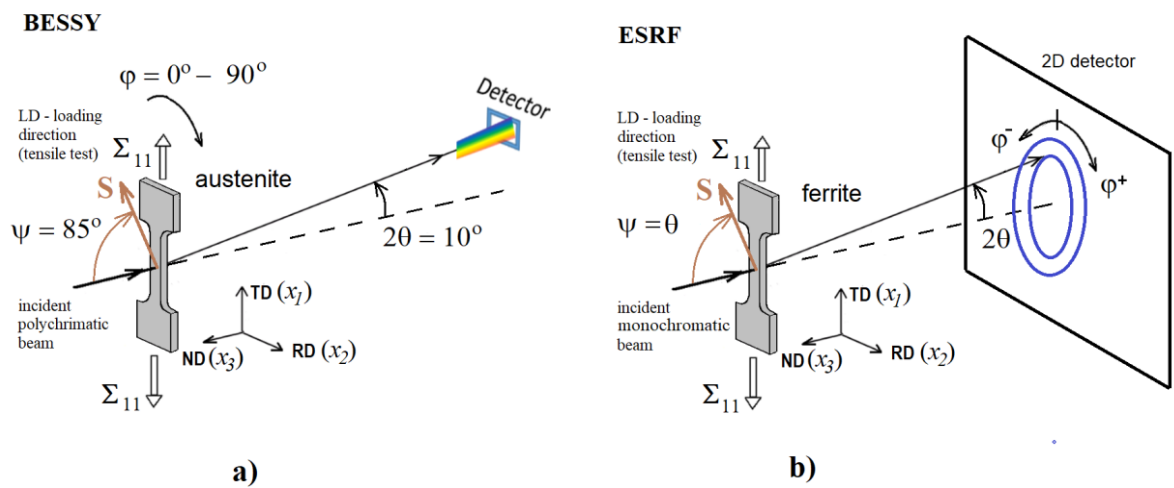

Fig. 3. An experimental setup used for lattice strain measurement in the case of austenitic steel by ED diffraction (a) and ferritic steel by AD diffraction (b). In the case of ED method the measurements were done for positive angles $\varphi$, while the positive $\varphi^+$ and negative $\varphi^-$ are available from AD diffraction rings recorded by 2D detector. The stress tensors and orientation of the scattering vector are defined with respect to X coordinates for which $x_1 \parallel$ TD, $x_2 \parallel$ RD and $x_3 \parallel$ ND.

## 2.2.1 Energy-dispersive diffraction measurements

For austenitic steel, the stress measurements performed in situ during the tensile test were made using synchrotron ED diffraction at BESSY (EDDI@BESSYII beamline, HZB, Berlin) using a white beam (wavelength in the range λ: 0.18 - 0.3 Å) [54,55]. The primary beam cross-section was equal to 1 x 1 mm², and a double slit system restricted the angular divergence in the diffracted beam with apertures of 0.1 x 5 mm² to $\Delta\theta \leq 0.005°$ (Fig. 3a).

Gathered diffraction line profiles and calculations of the lattice strains for various scattering vector orientations were used to determine the stresses. Diffractograms were collected with the steps of 0.1 vs. $cos^2\varphi$ (Fig. 3), within the range of $\varphi = (0°, 90°)$, in symmetrical transmission mode for a constant $2\theta = 10°$ scattering angle. Note that in the presented experiments, the evolution of interplanar spacings or dependence of $F_{ij}(hkl, \psi, \varphi, f)$ are presented versus $cos^2\varphi$ instead of $sin^2\psi$, as usual. This is because of the specific geometry of the measurements in which the tilt of the scattering vector from normal to the sample direction is given by constant angle ψ. However, it can be easily shown that in the case of the quasi-isotropic sample with negligible second-order stresses, the $cos^2\varphi$ plots should be linear, and the deviations from linearity are caused by crystallographic texture or/and second-order stresses. In the present work, the measurement was carried out for the non-loaded sample (initial), next for specific tensile loads (characterized by $\Sigma_{11}$) applied to the sample and finally during elastic unloading of the load. To do this, a load rig (from Walter + Bai AG) with a maximum load of 20 kN mounted on a Newport quarter circle cradle segment was employed for the in situ mechanical test. During plastic deformation, diffraction data were collected for fixed sample strain ($E_{11}$) after stabilization of the load applied to the sample.

Diffraction peaks were fitted with the pseudo-Voigt function to determine their positions $E_{hkl}$ versus energy scale. The interplanar spacings $<d>_{hkl}$ were evaluated using the following equation:

$$<d>_{hkl} = \frac{hc}{2sin\theta} \frac{1}{E_{hkl}} \tag{8}$$

where: $c$ - speed of light and $h$ -Planck constant.

As a result, the interplanar spacings for many diffraction *hkl* lines were simultaneously measured for given values of $cos^2\varphi$. Such measurement enables to determine $\sigma_{11}^I$ component of the stress tensor in the direction of the applied stress $\Sigma_{11}$.

### 2.2.2 Angle-dispersive diffraction

For ferritic steel, the in situ stress measurements were performed during the tensile test using AD diffraction at the ID15 synchrotron beamline (ESRF, Grenoble, France). The applied high-energy synchrotron radiation with wavelength λ = 0.14256 Å and a beam size of 100 μm x 100 μm enabled transmission measurements in the interior of the samples having a square cross-section with a side length of 1.5 mm (Fig. 3b). A square CCD detector (Thales PIXIUM 4700) was used to capture two-dimensional diffraction patterns during 10-second exposures separated by 5-second intervals. It was possible to conduct diffraction measurements in situ throughout a continuous tensile test due to the short data collection period.

The Fit2D software [56] was used to handle the collected data by the integration of 2D sectors with an angular size equal to $\Delta\varphi$ = 2° and converting them into the 1D ones composed of intensity dependence vs. *2θ* scattering angle. The theoretical functions were then fitted to the 1D diffractograms using Multifit software [57]. The positions of the diffraction peaks were found by adjusting the pseudo-Voigt function vs. *2θ*, and the interplanar spacings $<d>_{hkl}$ for different *{hkl}* planes were determined from the Bragg law:

$$<d>_{hkl} = \frac{\lambda}{2sin\theta_{hkl}} \tag{9}$$

Similarly, as in the previous experiment (ED diffraction for austenitic sample) the $<d>_{hkl}$ spacings can be determined simultaneously for many reflections *hkl*, for given values of $cos^2\varphi$

(with a small step of $\Delta\varphi = 2°$). Therefore, the $\sigma_{11}^I$ component of stress tensor (in the direction of the applied load Fig. 3b) can be determined with small increments $\Delta\Sigma_{11}$ corresponding to 10-second exposures separated by 5-second intervals during the continuous tensile test.

It should be emphasized that the performed experiments were done using both ED and AD techniques. This allows to verify whether the proposed second-order stress testing methodology can be applied to these techniques and to ensure that the observed phenomena are actually caused by the processes occurring in the material, and not the artefacts leading to changes in the 2θ angle in the AD method or energy shifts in the ED method. An important question concerning the performed experiment is whether enough information about second order stresses can be obtained and whether the experimental data are representative for the studied samples. To check this, the analysis of grains' contribution to the diffracted beam and the obtained results was done. At first, the integration paths in the Euler space were found for each experimental point, and the contribution of the grains to diffracted beam intensity was calculated considering the ODF function. This procedure of integration path determination is the same as when the pole figures from ODF are calculated [50]. Then the measure of orientations' contribution to the diffraction peak was determined, taking into account the ODF values along the determined path. The traces in Euler space were found for all experimental points, and the contribution measure of grains was summed over all measuring points. The so-obtained function of orientation contribution was normalized in the same way as ODF [50] and presented in Fig. 4 for both studied materials. As seen in Fig. 4 a, the experimental information is obtained mostly for preferred orientations (which contribution is dominating - cf. Fig. 2a), and on the contribution function, irregular spots corresponding to more informative regions are seen. This is the effect of the distribution of integration paths in the Euler space. In the case of ferritic steel, the step of $\varphi$ angle was very small (step of $\Delta\varphi = 2°$), and as a result, 46 points were obtained for each $< a(\psi, \varphi) >_{hkl}$ vs. $cos^2\varphi$ plot, while in the case of austenitic sample only

11 points were measured for each plot. Therefore, the orientation contribution function is very smooth and it is very similar to the ODF presented in Fig. 2b; however, it is seen that the regions with a small value of Φ - Euler angle are more representative. It can be concluded that for both tested samples, the preferred orientations contribute the most to the experiment; however, information from weak orientations is also included in the results. It is also seen that the informative region is distributed over a large part of the Euler space; therefore, the obtained results are representative for the majority of grains in the studied samples.

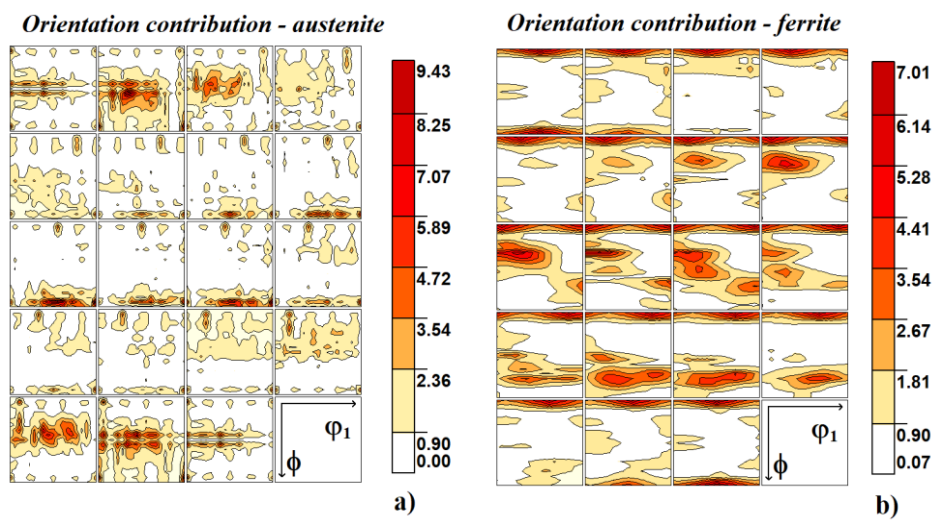

Fig. 4. The normalized orientation contribution function calculated for the integration paths corresponding to experimentally measured diffraction peaks with the weights given by ODFs for (a) austenitic and (b) ferritic samples. The levels express multiples-of-equal-contributions.

3. Results

**3.1 Validation of the XSF**

In order to perform a correct stress analysis, it is necessary to appropriately determine the elastic diffraction constants, especially for elastically anisotropic crystallites. Thus, in the first step, the theoretically predicted constants using the grain interaction models (Reuss, Voigt, Kröner) and the crystallographic texture were compared with the experimental results. Elastic constants of austenite and ferrite and anisotropy of elastoplastic deformation were verified by analyzing the lattice strains measured for different *hkl* reflections. To do this, the interplanar spacings were measured before and after a considerable change in the applied load (corresponding to increment $\Delta\Sigma_{11}$). The *F$_{11}$* constants were calculated from Eq. 2. In the case of austenite, the sample was subjected to elastoplastic deformation up to the applied stress $\Sigma_{11}^{(1)}$ = 360 MPa and then completely unloaded to $\Sigma_{11}^{(2)}$ = 0 MPa (i.e. $\Delta\Sigma_{11}$ = -360 MPa). As shown in Fig. 2a, the texture did not change significantly during the tensile test; therefore, it was insignificant in the calculation of the XSF. The ferritic sample broke after the last diffraction measurement at 5% strain because, in the experiment, a stress control mode was used with a constant step of the applied load (due to the low hardening of the ferritic sample, the strain increased sharply in the last step of increasing load). Thus, the *F$_{11}$* factors were calculated based on the experimental data for sample loading from a small load applied to fix the sample $\Sigma_{11}^{(1)}$ = 5 MPa up to $\Sigma_{11}^{(2)}$ = 352MPa (i.e., $\Delta\Sigma_{11}$ = 347MPa). Model calculations were performed considering the initial texture shown in Fig. 2b. The calculated and experimental results of *F$_{11}$* vs. $cos^2\varphi$ are presented in Figs. 5 and 6.

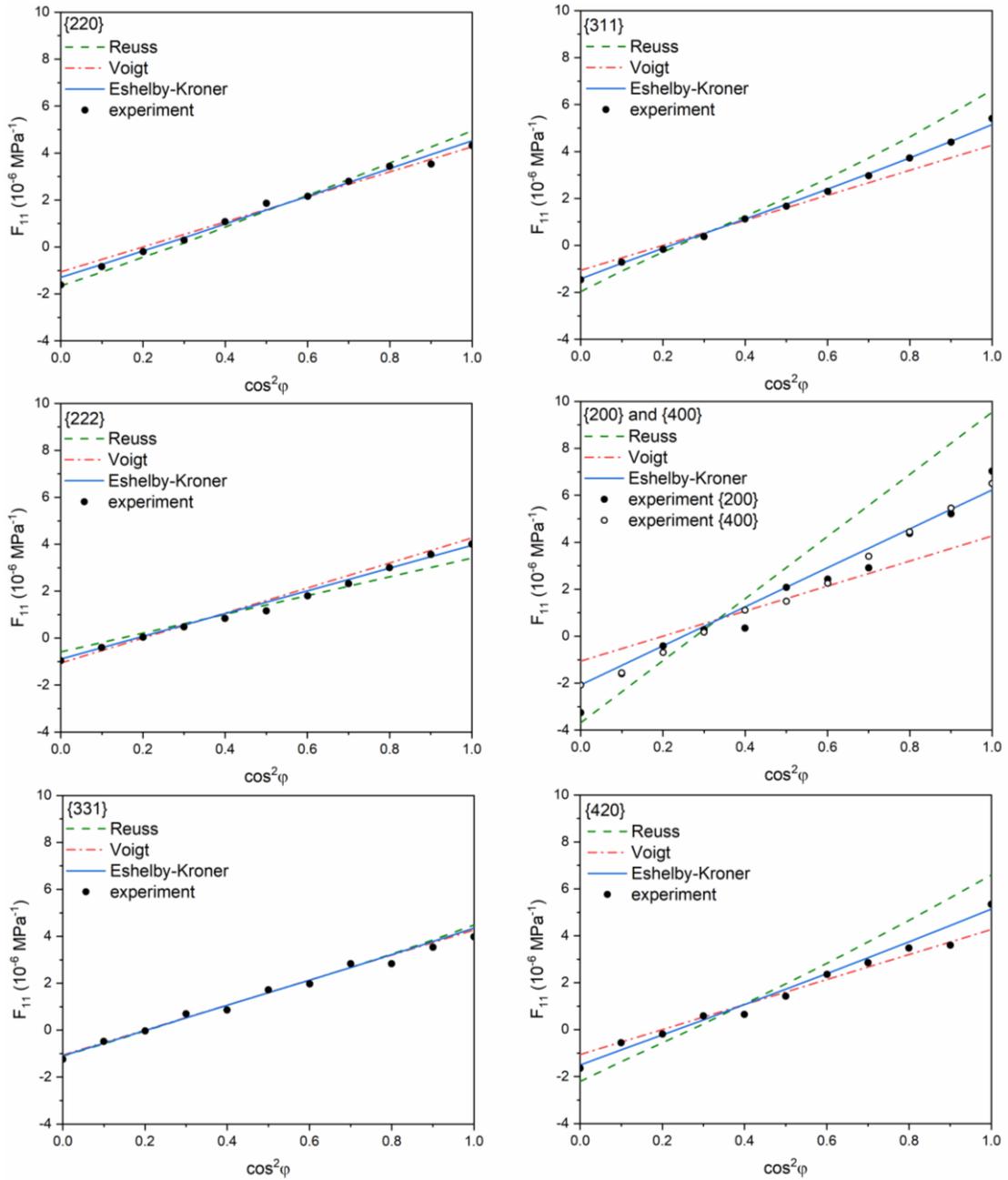

Fig. 5. $F_{11}$ vs. $\cos^2\varphi$ curves for experimentally obtained data for austenitic steel, using ED diffraction. Six different reflecting planes were used together with the following grain elastic interaction models: Reuss, Voigt, and Eshelby-Kröner. The experimental points (dots) correspond to the stress increment $\Delta\Sigma_{11} = -360$ MPa, during sample unloading.

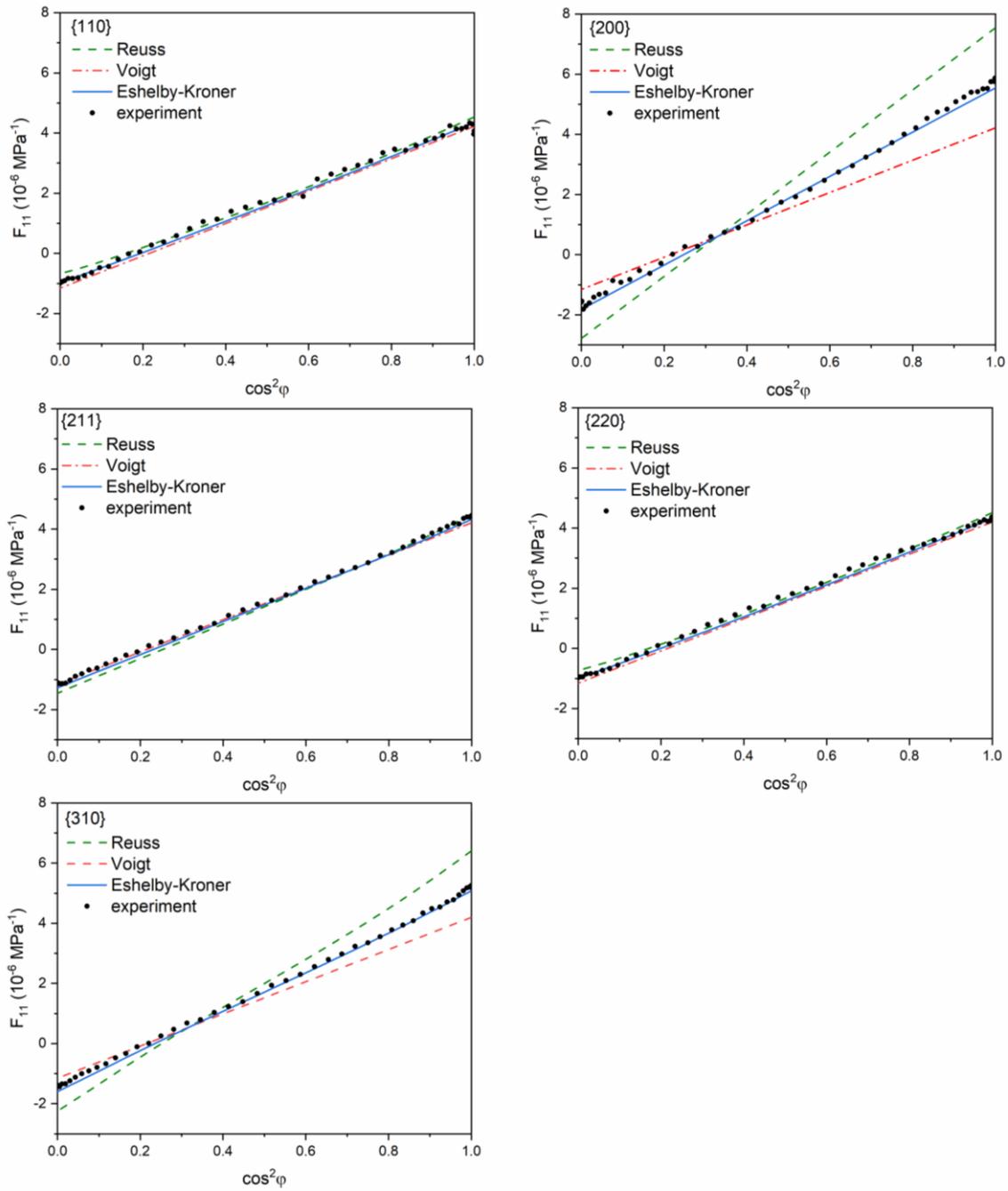

Fig. 6. $F_{11}$ vs. $\cos^2\varphi$ curves for experimentally obtained data for ferritic steel, using AD diffraction. Five different reflecting planes were used together with the following grain elastic interaction models: Reuss, Voigt, and Eshelby-Kröner. The experimental points (dots) correspond to the stress increment $\Delta\Sigma_{11} = 347.3$ MPa during elastic loading.

In Figs. 5 and 6, it can be clearly seen that the values of $F_{11}$ (XSF) calculated by the Eshelby-Kröner model fit best the experimental results for both samples. Therefore, the stress measured by X-ray diffraction was determined using the XSF derived by the Eshelby-Kröner model with the ODF function shown in Fig. 2. It should be noted that despite crystallographic texture the non-linearities of the $F_{11}$ vs. vs. $cos^2\varphi$ curves are very small, at least for the chosen direction of lattice strains measurement. This is confirmed both by the experimental and theoretical results.

It should be emphasized that the in situ test of factors $F_{11}$ performed with multiple *hkl* reflections is very rigorous because many groups of crystallites with different orientations are involved in the diffraction experiment (cf. orientation contribution functions shown in Fig. 4). It proves that the used single crystal elastic constants and grain interactions are correctly approximated.

It is worth noting that, in our work, the careful verification of the applicability of the model for calculation XSFs is necessary because to evaluate the effect of second-order stresses, at first the effect of the stress applied to the sample must be known. It is well known that for textured samples, the nonlinearities of the $<a(\psi,\varphi)>_{hkl}$ vs. $cos^2\varphi$ (or $<a(\psi,\varphi)>_{hkl}$ vs. $sin^2\psi$ in standard stress analysis) can be caused by nonlinearities of $F_{11}$ vs. $cos^2\varphi$ dependence (for the uniaxial test). Thus to minimize such an effect, the orientation of the scattering vector was changed between RD and TD because, in this case, the nonlinearities on $F_{11}$ vs. vs. $cos^2\varphi$ curves were small, as shown in Figs. 5 and 6. Therefore we avoided the problem of overlapping the nonlinearities coming from the anisotropy of XSF (due to texture) and those caused by the second-order stresses.

## 3.2 Stress evolution during elastoplastic deformation

Next, the results of stress measurements during the controlled tensile test were analyzed for both samples using the verified XSF. The interplanar spacings (Eqs. 8 and 9) were measured in situ for each certain number of loads in the elastic and plastic deformation range during loading and unloading. As already mentioned, the ferritic sample fractured during the test before the unloading step. The force was applied along TD for both samples. Applying Eq. 3 and the fitting procedure for the experimental data, the set of quantities: first-order stress component $\sigma_{11}^I$, $a_0$ and the $q$-factor were determined from Eq. 5 using the least square procedure. In fitting all lattice parameters $<a(\psi,\varphi)>_{hkl}^{\sigma}$ measured for different $hkl$ reflections were used simultaneously for a given load or unloaded sample. The $\sigma_{22}^I$, $\sigma_{33}^I$ and shear first-order stress components were assumed equal to zero because uniaxial stress was imposed during the tensile test. The measurement of the $<a(\psi,\varphi)>_{hkl}^{\sigma}$ for orientations of scattering vector changed between RD ($x_2$) and TD ($x_1$) is enough to determine $\sigma_{11}^I$ and $a_0$, similarly as in standard X-ray measurement with the assumption of zero normal stress ($\sigma_{33}^I = 0$). It is worth noting that,, the adjustment of the $q$-factor is based on model-calculated lattice strains $<\tilde{\varepsilon}(\psi,\varphi)>_{hkl}^{pi}$ fitted to experimental nonlinearities of the $<a(\psi,\varphi)>_{hkl}^{\sigma}$ vs. $cos^2\varphi$ plots. The lattice strains $<\tilde{\varepsilon}(\psi,\varphi)>_{hkl}^{pi}$ are caused by all stress components of second-order stresses $\tilde{\sigma}_{ij}^{II,pi}$, and information about the variations of these components and the form of the stress tensor for individual grains is determined by the model. The factor $q$ multiplies the magnitude of lattice strains $<\tilde{\varepsilon}(\psi,\varphi)>_{hkl}^{pi}$ and can be used as the scaling factor for all components of the model stress tensor $\tilde{\sigma}_{ij}^{II,pi}$, to find their magnitudes in the real sample $\sigma_{ij}^{II,pi}$ (for all grains simultaneously). Therefore, knowing the values of the $q$-factor, the second-order stresses $\sigma_{ij}^{II,pi}$ (full tensor) were calculated from Eq. 6 for each orientation of the grain lattice.

The self-consistent model of elastoplastic deformation elaborated by Lipinski and Berveiller [44] was used to calculate the values $<\tilde{\varepsilon}(\psi,\varphi)>_{hkl}^{pi}$ needed to interpret the experimental results and to determine the second-order stresses from Eq. 6. The input file generated for EPSC model, contained 10000 spherical inclusions (representing grains) with equal volume fraction and distribution of orientations determined based on initial experimental ODF (Fig. 2). Single crystal elastic constants defined with respect to the crystal lattice and the residual stresses $\sigma_{11}^{I,res}$ measured for the initially non-loaded sample were assigned to each inclusion. The input files created in this way were used to simulate the elastic-plastic tensile deformation of ferritic and austenitic samples using the EPSC model.

In calculations, the isotropic hardening matrix and the linear hardening law were assumed. All slip systems in all grains (for the given sample) had the same initial Critical Resolved Shear Stress (CRSS). The parameters characterizing slip system activation (CRSS) and hardening (*H*) [44] were optimized to adjust the theoretical macroscopic stress-strain curves to the experimental ones. The values of optimal parameters for both samples are given in Table 3, while the fitted (model) and experimental macroscopic mechanical (obtained during in situ tensile test) curves will be presented in Figs. 11 and 12, together with the plots obtained from diffraction.

Previously obtained values of CRSS and *H* parameters optimized by fitting the EPSC model to an experimental mechanical stress-strain plot for the austenitic sample are given by Neil at al. [40], and they are not far from those given in Table 2, i.e., $\tau_c^0$ = 93 MPa and *H* = 375 MPa (represented by *θ₀* in that work). Certainly, these values are not the same because they depend on the material microstructure, chemical composition and performed treatment. Concerning studied ferritic steel, we have not found analogous results, but it is well known that $\tau_c^0$ is related to the yield stress which could be very different for different types of microstructure and material processing. However, a very small work hardening ($H\rightarrow 0$, at the beginning of plastic deformation, c.f. Table 3) is characteristic for ferrite, as was found, e.g., in pearlitic steel [32].

Table 3. Input parameters characterizing the initial microstructure of the investigated materials

| Structure | Slip systems | Initial CRSS $\tau_c^0$ (MPa) | Hardening parameter H (MPa) |
|---|---|---|---|
| bcc (ferrite) | <111>{1-10} <111>{11-2} <111>{12-3} | 190 | 0 |
| fcc (austenite) | <110>{111} | 84 | 270 |

In the stress analysis using the least square method, seven reflections: 200, 220, 222, 311, 331, 400, 420, and five reflections: 110, 200, 211, 310, 220 were considered for austenitic and ferritic steel, respectively. The analysis was carried out for the initial non-loaded samples and for the samples subjected to various uniaxial loads applied during in situ measurements. Finally, the stresses were determined in the unloaded samples. In the case of the fractured ferritic specimen, the measurements were performed in a different location than during the tensile test to avoid the effect of stress heterogeneity close to the fracture surface.

The first example of the $<a(\psi,\varphi)>_{hkl}$ vs. $cos^2\varphi$ curves is shown for the initial (non-loaded) samples in Fig. 7. The theoretical plots are presented for two assumptions, i.e. when the plastic incompatibility stresses are not present (dashed lines, $q=0$) and when their influence is taken into account. The $q$ parameter is determined from Eq. 5 (continuous lines).

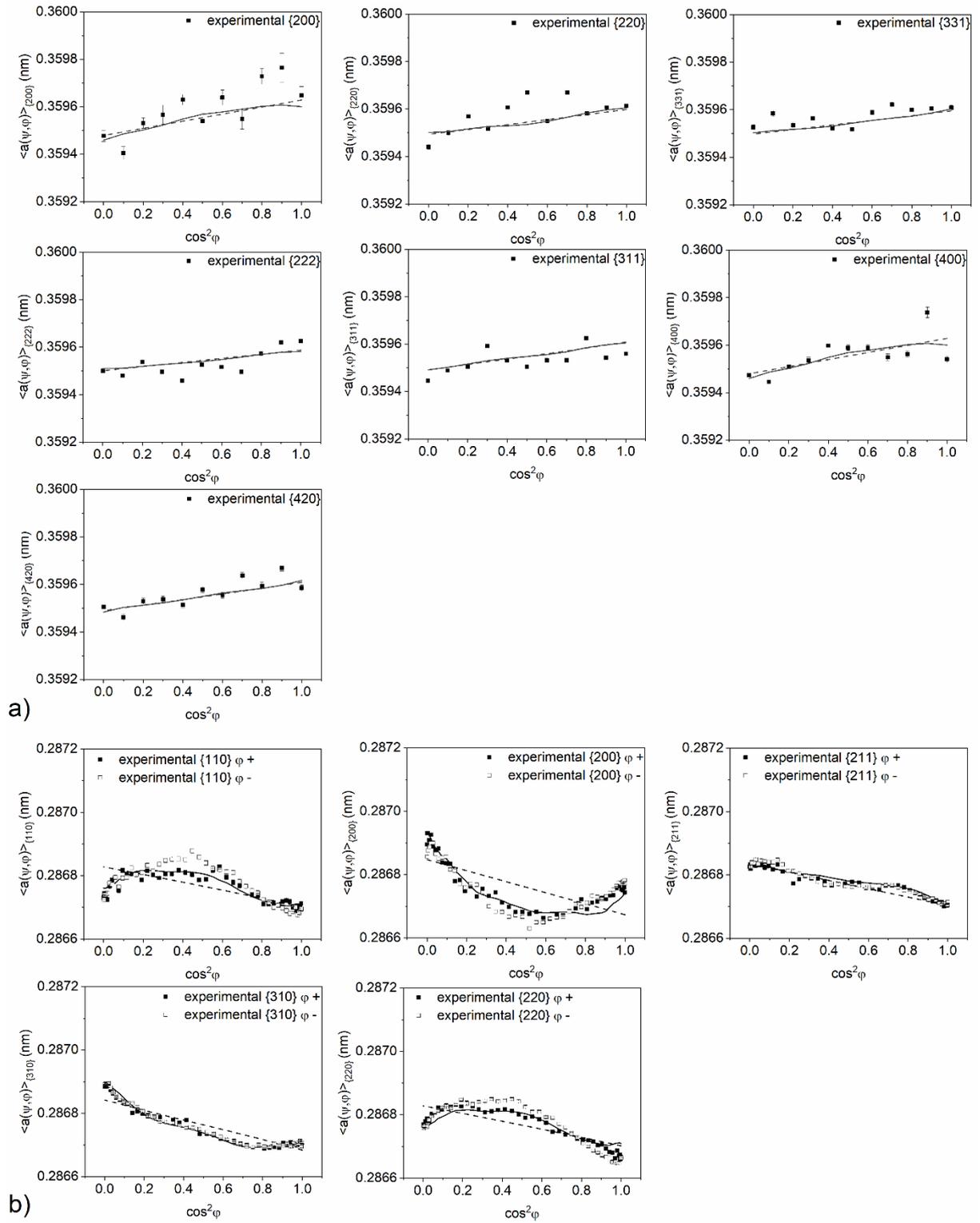

Fig. 7. Comparison of the experimental data (points) with the calculated $<a(\psi,\varphi)>_{hkl}$ vs. $cos^2\varphi$ plots obtained for adjusted *q*-parameter (solid line) and the *q* = 0 (dashed line). The results are shown for the initial non-loaded (a) austenitic sample ($\Sigma_{11}$ = 0 MPa; both lines, solid and dashed, overlap each other) and (b) ferritic sample ($\Sigma_{11}$ = 5 MPa, the small load was applied to fix the sample).

In the case of initial austenitic sample, the $q$-fitting method had no impact on the results. Thus the small non-linearities in the experimental $<a(\psi,\varphi)>_{hkl}$ vs. $cos^2\varphi$ data do not coincide with those simulated for the uniaxial tensile elastoplastic deformation. It means that the residual second-order stresses generated during sample preparation, cannot be determined for the initial sample, and only the value $\sigma_{11}^I$ of tensile residual first-order stress in the measured gauge volume was found. On the contrary, when second-order stress is taken into account in stress estimation for the initial ferritic sample (with adjusted $q$ parameter), the results of the least square fitting are much improved, as shown in Fig. 7 b. This means that the residual second-order stresses in the initial state of the ferritic sample well coincide with those predicted by the EPSC model for plastic deformation occurring during tensile test. Moreover, the first-order compressive stress $\sigma_{11}^I$ was obtained for the measured gauge volume. This sample contains residual stresses due to the cold rolling procedure followed by sample preparation.

The second example of the $<a(\psi,\varphi)>_{hkl}$ vs. $cos^2\varphi$ curves is shown for the loads applied in situ during purely elastic tensile deformation (Fig. 8). In this case, the slopes of the plots significantly changed, showing the tensile character of the first-order stresses for both studied samples. However, the analysis of the second-order stresses shows the same results as for initial samples, i.e., no correlation between model and experiment non-linearities in the case of the austenitic sample (adjustment of $q$ parameter does not improve the quality of fitting) and very good correlation of non-linearities in the case of the ferritic sample, as shown in Fig. 8 a and 8 b, respectively. That means that the state of second-order stresses did not change significantly during the elastic loading of both samples.

The next example of the $<a(\psi,\varphi)>_{hkl}$ vs. $cos^2\varphi$ curves was chosen for the significant plastic deformation of both samples under applied load (Fig. 9). In this case, the slopes of the plots increased, indicating the increase of applied tensile load. Also, interesting evolution of non-linearities was observed in the case of the austenitic sample (Fig. 9 a). Indeed, the solid lines representing results with adjustment of the $q$ parameter started to match the experimental points. It implies that the non-linearities of the $<a(\psi,\varphi)>_{hkl}$ vs. $cos^2\varphi$ plots predicted by the model more closely match those obtained from measurements. Therefore the plastic incompatibility second-order stresses (i.e., the second-order stresses induced by plastic deformation) can be determined using Eq. 6. It can also be concluded that the stress state for the grains was transformed entirely due to the plastic deformation, i.e., second-order stresses occurring initially in the prepared sample have been completely replaced with stresses generated due to plastic deformation during the tensile test. In the case of a ferritic sample, similarly as in the initial sample, the non-linearities in the sample subjected to plastic deformation are still well reproduced by model data (the fitting of the solid line is unquestionably better than that of the dashed line, as shown in Fig. 9 b. It means that no significant modification of the second-order stresses occurred due to plastic deformation during the tensile test.

Finally, the last example of the $<a(\psi,\varphi)>_{hkl}$ vs. $cos^2\varphi$ curves referring to the results for the unloaded austenitic sample and the fractured ferritic sample is presented in Fig. 10. When comparing the initial (Fig. 7 a) and unloaded state (Fig. 10 a) for the austenitic sample, it can be seen that the dependence $<a(\psi,\varphi)>_{hkl}$ vs. $cos^2\varphi$ changed significantly. This proves that the state of residual second-order plastic incompatibility stress was changed entirely during the plastic deformation of the sample. The essential improvement of fitting quality for the unloaded sample, when the $q$ parameter is adjusted, confirms that the analysis is carried out correctly when the model data from the EPSC model for the tensile test are used. The results obtained for the fractured ferritic sample (Fig. 10 b) compared with the initial state (Fig. 7 b) show that no significant change occurred in the state of second-order stresses.

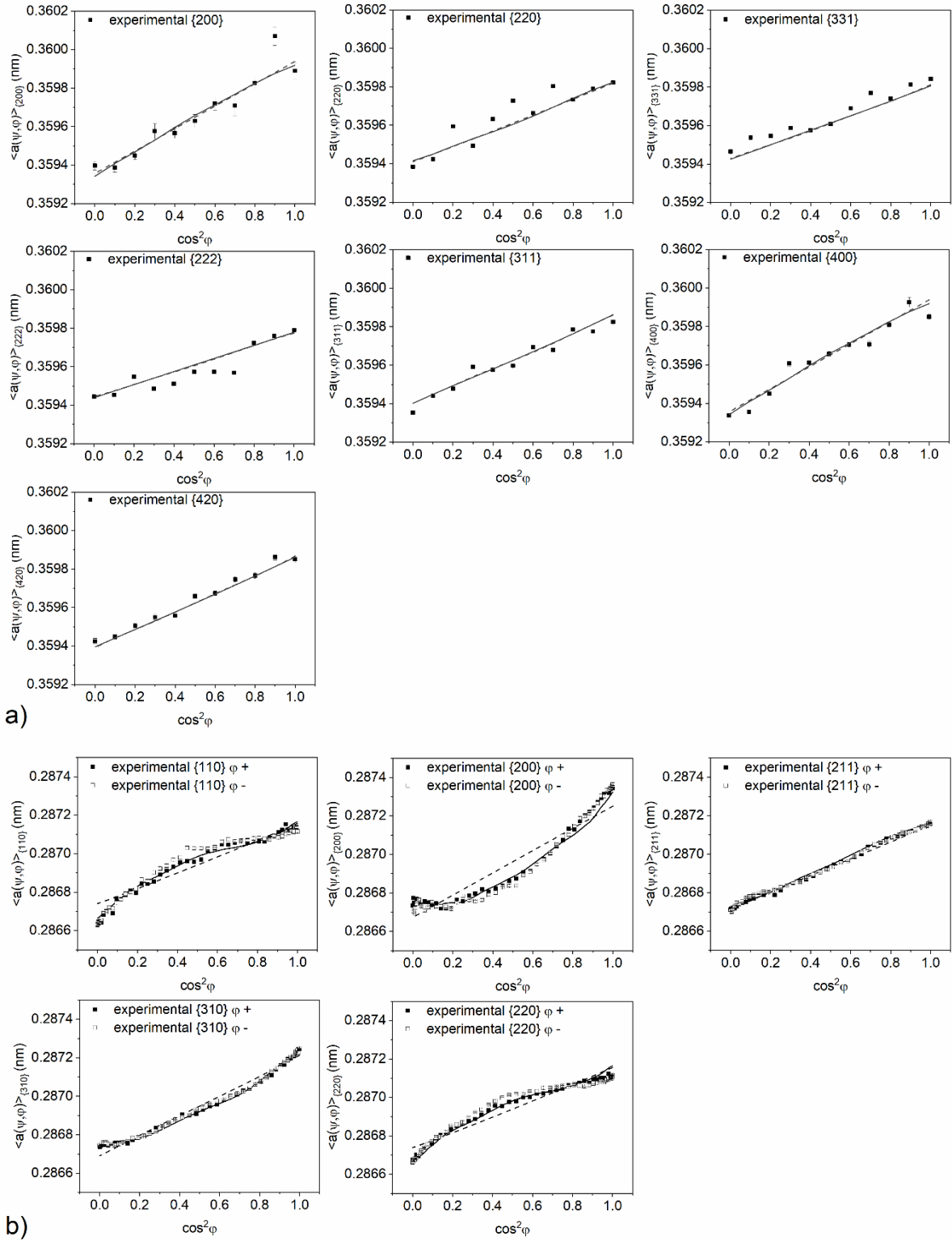

Fig. 8. Comparison of the experimental data (points) with the calculated $<a(\psi,\varphi)>_{hkl}$ vs. $cos^2\varphi$ plots obtained for adjusted $q$-parameter (solid line) and the $q = 0$ (dashed line). The results are shown for the elastic range of deformation for (a) austenitic sample ($\Sigma_{11}$ = 140 MPa) and (b) ferritic sample ($\Sigma_{11}$ = 352 MPa). The solid and dashed lines overlap each other in the case of austenite.

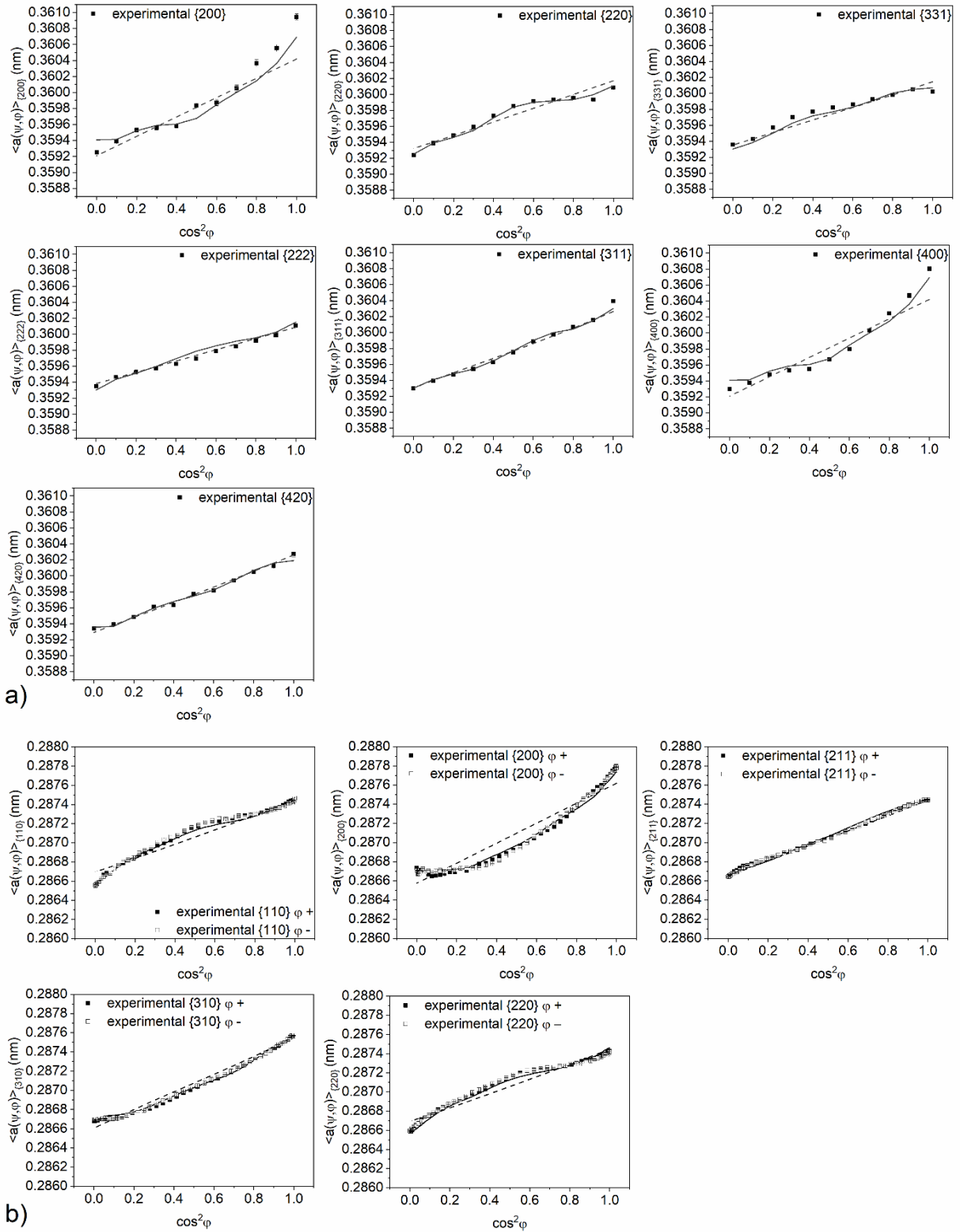

Fig. 9. Comparison of the experimental data (points) with the calculated $<a(\psi,\varphi)>_{hkl}$ vs. $cos^2\varphi$ plots obtained for adjusted $q$-parameter (solid line) and the $q = 0$ (dashed line). The results are shown for the load causing plastic deformation for (a) austenitic sample ($\Sigma_{11}$ = 360 MPa) and (b) ferritic sample ($\Sigma_{11}$ = 585 MPa).

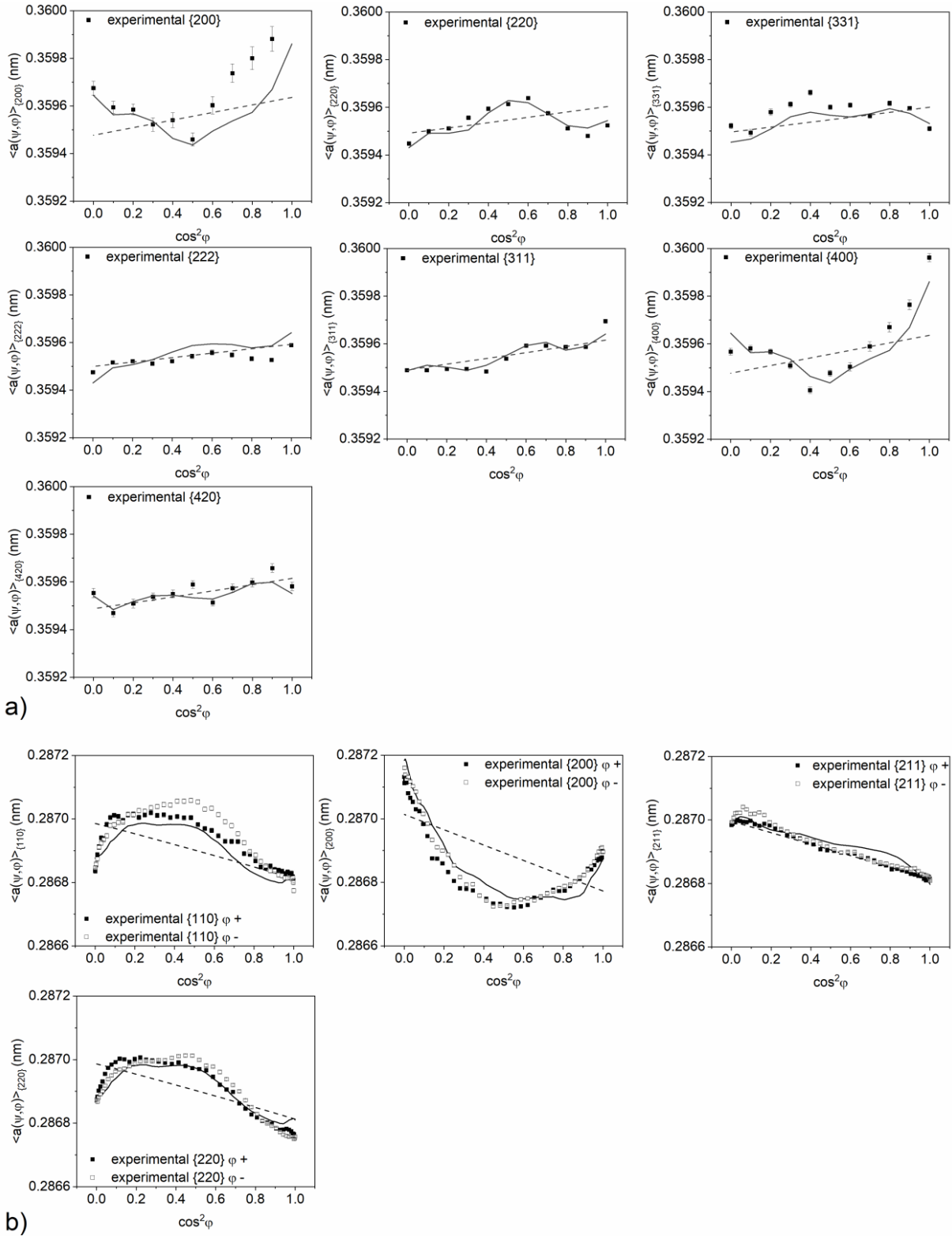

Fig. 10. Comparison of the experimental data (points) with the calculated $< a(\psi, \varphi) >_{hkl}$ vs. $cos^2\varphi$ plots obtained for adjusted $q$-parameter (solid line) and the $q = 0$ (dashed line). The results are shown for the (a) unloaded austenitic sample ($\Sigma_{11} = 0$ MPa) and fractured ferritic sample ($\Sigma_{11} = 0$ MPa).

By applying Eq. 5 and by fitting the results from the model to the experimental data, the values of first-order stress $\sigma_{11}^{I}$ and the *q* factor could be found. Then the values of second-order stress $\sigma_{ij}^{II,pl}$ for each polycrystalline grain was determined using Eq. 6 in which the model stresses ase multiplied by *q* factor. Finally, the mean value of von Mises stress $\overline{\sigma_{Mises}^{II,pl}}$ over all grains was calculated for the initial, loaded, and unloaded/fractured samples (for details, see [8,9]). The so defined $\overline{\sigma_{Mises}^{II,pl}}$ is a good measure of mean magnitude of second-order plastic incompatibility stresses because the hydrostatic stress computed from $\sigma_{ij}^{II,pl}$ is equal to zero for purely plastic deformation, as it was also verified using EPSC model. In Figs. 11 a and 12a, the so-obtained values of stresses $\sigma_{11}^{I}$ and $\overline{\sigma_{Mises}^{II,pl}}$ are presented vs. sample strain *E₁₁*. To determine the evolution of the stress state during the loading process, the first- and second-order stress are presented as the function of the superposition of the imposed stress ($\Sigma_{11}$) and residual stress in the initial non-loaded sample ($\sigma_{11}^{I,res}$), i.e.: $\Sigma_{11} + \sigma_{11}^{I,res}$ (Figs. 11 b and 12 b). These graphs illustrate the comparison of the first-order stress evolution in the irradiated volume during sample loading determined in two ways: as the sum of the residual stress ($\sigma_{11}^{I,res}$) superposed with the imposed stress ($\Sigma_{11}$) and as the stress determined directly from the diffraction experiment ($\sigma_{11}^{I}$) for the corresponding load. Certainly, the two values should be equal if the sample and stress state is homogenous. The difference between them indicates heterogeneity of the stress distribution across the sample along the direction *x₂* (Fig. 3) caused by the material processing and preparation of the dog-bone-shaped samples. It should be emphasized that due to the much smaller size of the synchrotron beam spot compared to the sample width (in direction *x₂*), the local stress state in the center of the samples was determined in both measured initial samples. As mentioned before in the case of the experiment performed for ferritic sample (the high energy synchrotron radiation with wavelength λ = 0,14256 Å at ESRF) the beam size of 100 µm x 100 µm enabled transmission measurements in the interior of the sample having 1.5 mm in width. The primary beam cross-section for the EDDI experiment using a white X-ray beam was equal to 1 x 1 mm2, and again the spot size was smaller than the sample's width equal to 5 mm.

The evolution of the second-order stresses $\overline{\sigma_{Mises}^{II,pl}}$ vs. ($\sigma_{11}^{I,res} + \Sigma_{11}$) are also shown in Figs. 11 b and 12b. In addition, in Figs. 11 and 12, the results of the EPSC prediction corresponding to the experimental data are presented (solid lines). The prediction starts from the initial small residual stress $\sigma_{11}^{I,res}$ (tensile for austenitic sample seen in Fig. 11; and compressive for ferritic sample seen in Fig. 12) and ends at the residual stresses value remained after the tensile test, so after sample unloading. It should be emphasized that the parameters of the model (Table 3) were adjusted to reproduce the dependence of the first-order stress (in the information volume seen by diffraction) determined as $\sigma_{11}^{I,res} + \Sigma_{11}$ on the sample strain $E_{11}$ (squares shown in Figs. 11 a and 12 a).

In order to confirm the correctness of the chosen model (Eshelby-Kröner) for the calculation of the stress factor $F_{11}$ the experimental values $\sigma_{11}^{I,res} + \Sigma_{11}$ are compared with the first-order stress $\sigma_{11}^{I}$ determined by diffraction using Eshelby-Kröner, Voigt, and Reuss model (Figs. 11 a and 12 a). An excellent agreement was obtained for the Eshelby-Kröner model, while the use of both other models leads to a significant discrepancy between the stress calculated from the applied load ($\sigma_{11}^{I,res} + \Sigma_{11}$) and that measured by diffraction ($\sigma_{11}^{I}$) for both investigated materials. It supports the choice of the Eshelby-Kröner model for this study, which has already been verified in section 3.1. It should be emphasized that the perfect agreement between $\sigma_{11}^{I}$ and $\sigma_{11}^{I,res} + \Sigma_{11}$ stresses were obtained for the elastic range of deformation (see Figs. 11 b and 12 b), where the solid red line (obtained from the EPSC model) illustrates the equality $\sigma_{11}^{I} = \sigma_{11}^{I,res} + \Sigma_{11}$. Slight deviations of the diffraction results from the solid red line are observed for the plastic deformation range, especially for the austenitic sample. This effect can be explained by a modification of the first-order residual stress distribution (heterogeneity) along the $x_2$ axis, leading to different residual stress $\sigma_{11}^I$ in the measured volume after the tensile test compared to the initial value before loading (cf. Fig. 11 b, loading and unloading).

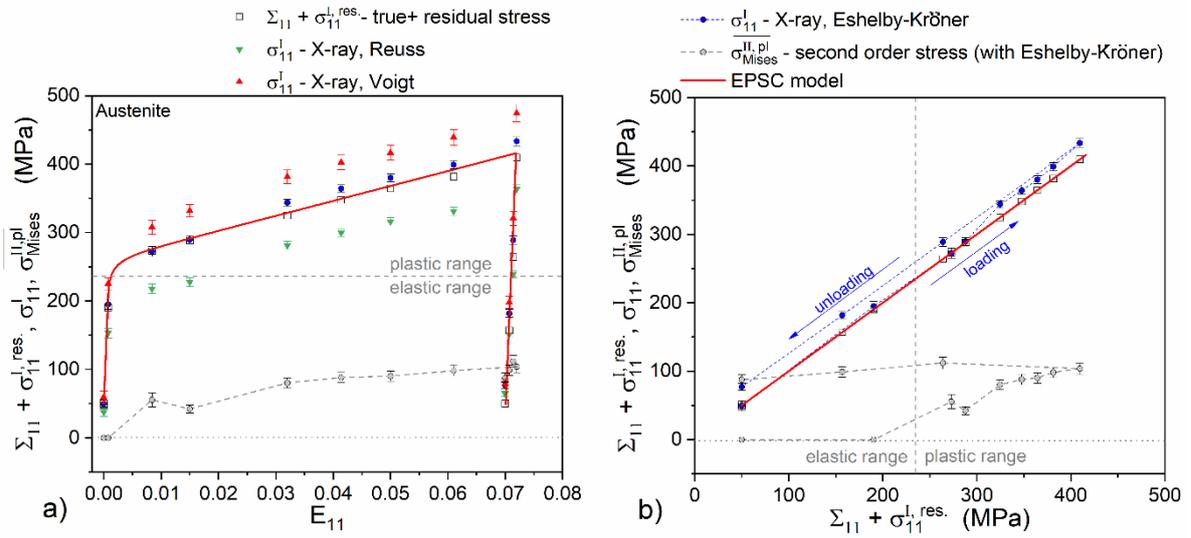

Fig. 11. Evolution of the experimental first-order stresses expressed in two ways: $\sigma_{11}^I$ and $\sigma_{11}^{I,res} + \Sigma_{11}$ compared with $\sigma_{11}^I$ predicted by the EPSC model, using parameters given in Table 3, for an austenitic sample subjected to tensile deformation. The evolution of determined second-order stresses mean von Mises stress ($\overline{\sigma_{Mises}^{II,pl}}$) is also presented. The plots vs. sample strain $E_{11}$ (a) and first-order stress $\Sigma_{11} + \sigma_{11}^{I,res}$ (b) are shown.

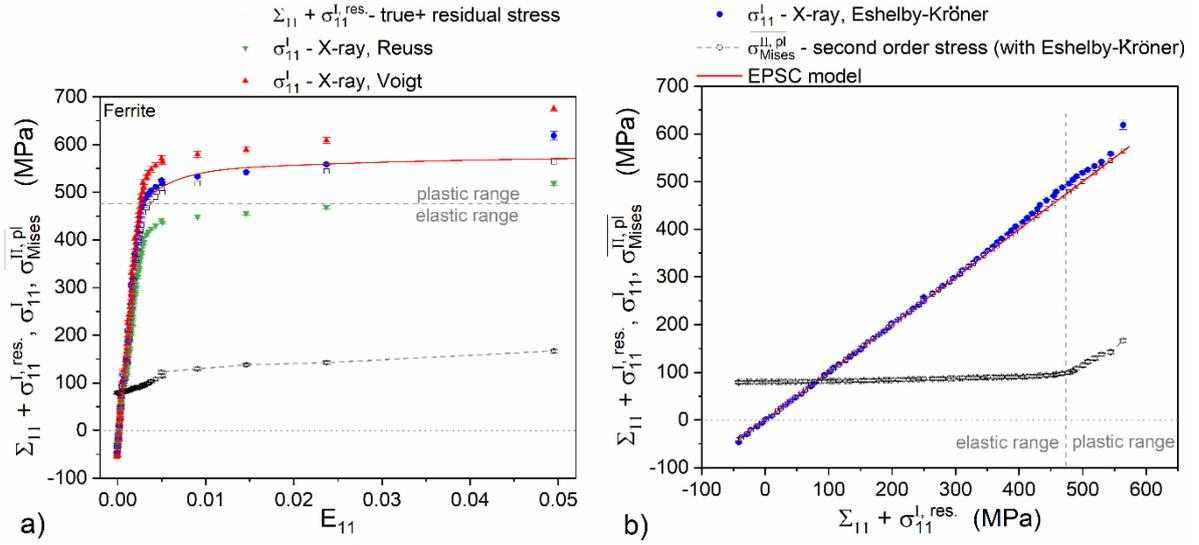

Fig. 12. The analogous evolutions as presented for the ferritic sample in Fig. 11.

Important results were obtained when the second-order stresses (characterized by $\overline{\sigma_{Mises}^{II,pl}}$) were analyzed. The evolution of $\sigma_{ij}^{II,pl}$ stresses should be discussed together with the changes in the figure of merit $\chi^2$ (Eq. 8) describing the least square fitting quality of the calculated lattice parameters $< a(\psi, \varphi) >_{hkl}$ to the experimental results.

In Fig. 13, the variation of $\chi^2$ vs. sample total strain $E_{11}$ is shown for two ways of data treatment. First assumes $q = 0$ (the influence of $\sigma_{ij}^{II,pl}$ is neglected), while the second takes into account the adjusted $q$-parameter. Both for austenitic (Fig. 13 a) and ferritic (Fig. 13 b) samples, the fitting is much better, i.e., $\chi^2$ is much lower when the $q$ is adjusted in Eq. 5 (excluding the two first points for the austenitic sample).

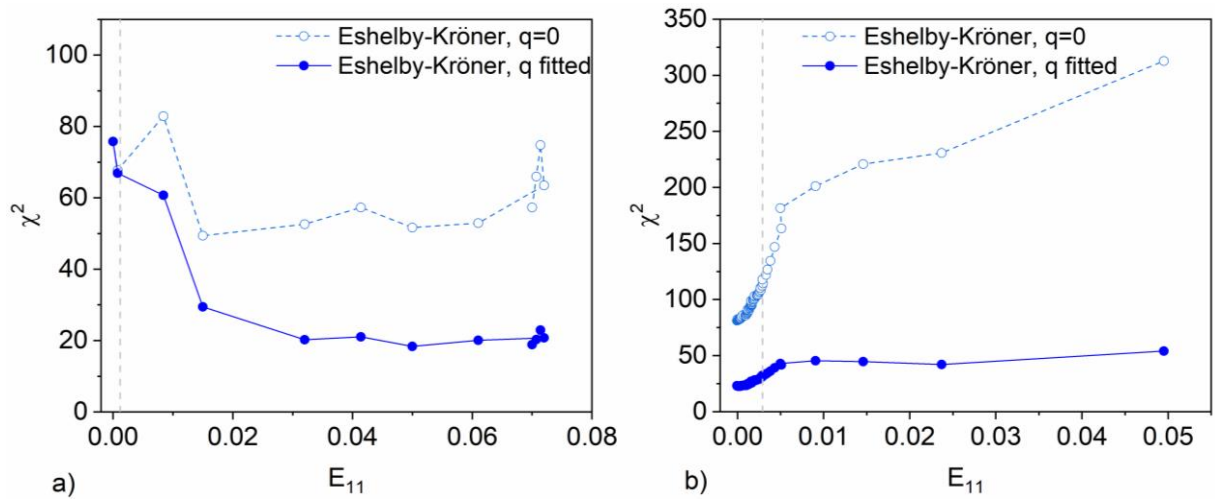

Fig. 13. Evolution of parameter $\chi^2$ in the function of sample strain $E_{11}$ for austenitic (a) and ferritic (b) samples subjected to a tensile test. A dashed vertical line separates the elastic range from the plastic deformation range.

In Fig. 14 the $a_0$ stress-free parameter obtained in stress analysis was shown for different deformation of the sample, using the same range for $a_0$ as for $<a(\psi,\varphi)>_{hkl}$ in Figs. 7 and 10. Comparing the variation of $<a(\psi,\varphi)>_{hkl}$ in Figs. 7 and 10 with the changes of $a_0$ in Fig. 14, it can be concluded that the value $a_0$ obtained from the analysis is almost constant during the tensile test, and their insignificant changes do not influence results concerning determined first- and second-order stresses.

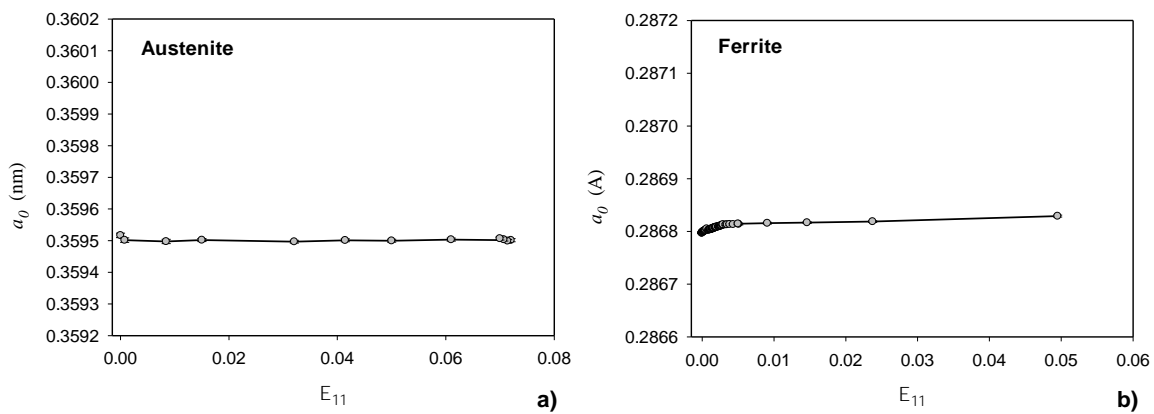

Fig. 14. The $a_0$ stress-free parameter determined for loaded and unloaded samples vs. sample strain. The same vertical scale as for the initial (Fig. 7) and unloaded/fractured samples was applied for better comparison (Fig. 10).

Finally, it is interesting to present the dependence of the plastic incompatibility second-order stresses on grain orientation (i.e. $\sigma_{ij}^{II,pl}$ obtained from Eq. 6 using our method). Because the full tensor $\sigma_{ij}^{II,pl}$ is built from 6 independent components, its presentation in Euler space is difficult. Therefore, the von Mises stress $\sigma_{Mises}^{II,pl}$ calculated from the tensor $\sigma_{ij}^{II,pl}$ for each orientation can be shown as the measure of the magnitude of the second-order plastic incompatibility stresses. The so calculated $\sigma_{Mises}^{II,pl}$ is shown in Euler space for austenitic (Fig. 15 a) and ferritic (Fig. 15 d) samples subjected to tensile deformation followed by unloading/fracture. It was found that there the minima of von Mises second-order stresses $\sigma_{Mises}^{II,pl}$ corresponds to maxima of ODF obtained by EPSC for austenitic steel (compare Fig. 15 a with Fig.15 b); however, no such correlations were found in the case of the ferritic sample. What is more, in some regions of Euler space, minima of $\sigma_{Mises}^{II,pl}$ correspond to minima of ODF – compare Fig. 15 d with Fig.15 e. Comparison of Figs. 15 b and e with Fig. 3 shows that the texture change during the performed deformation is very small for both studied samples.

Finally, the distribution of second-order stresses $\sigma_{Mises}^{II,pl}$ can be compared with the maximum value of the resolved shear stress RSS for all potentially active slip systems at given lattice orientations. It can be noticed that low values of the maximum RSS correspond to small values of von Mises second order plastic incompatibility stresses in the case of austenitic sample, c.f. Fig. 15 a and 15 c. In the case of ferritic samples, the correlation is similar but not so strong, i.e., for the low values of maximum RSS, the values of $\sigma_{Mises}^{II,pl}$ are also low, but not always small value of $\sigma_{Mises}^{II,pl}$ correspond to a low value of maximum RSS, c.f. Fig. 15 d and 15 f.

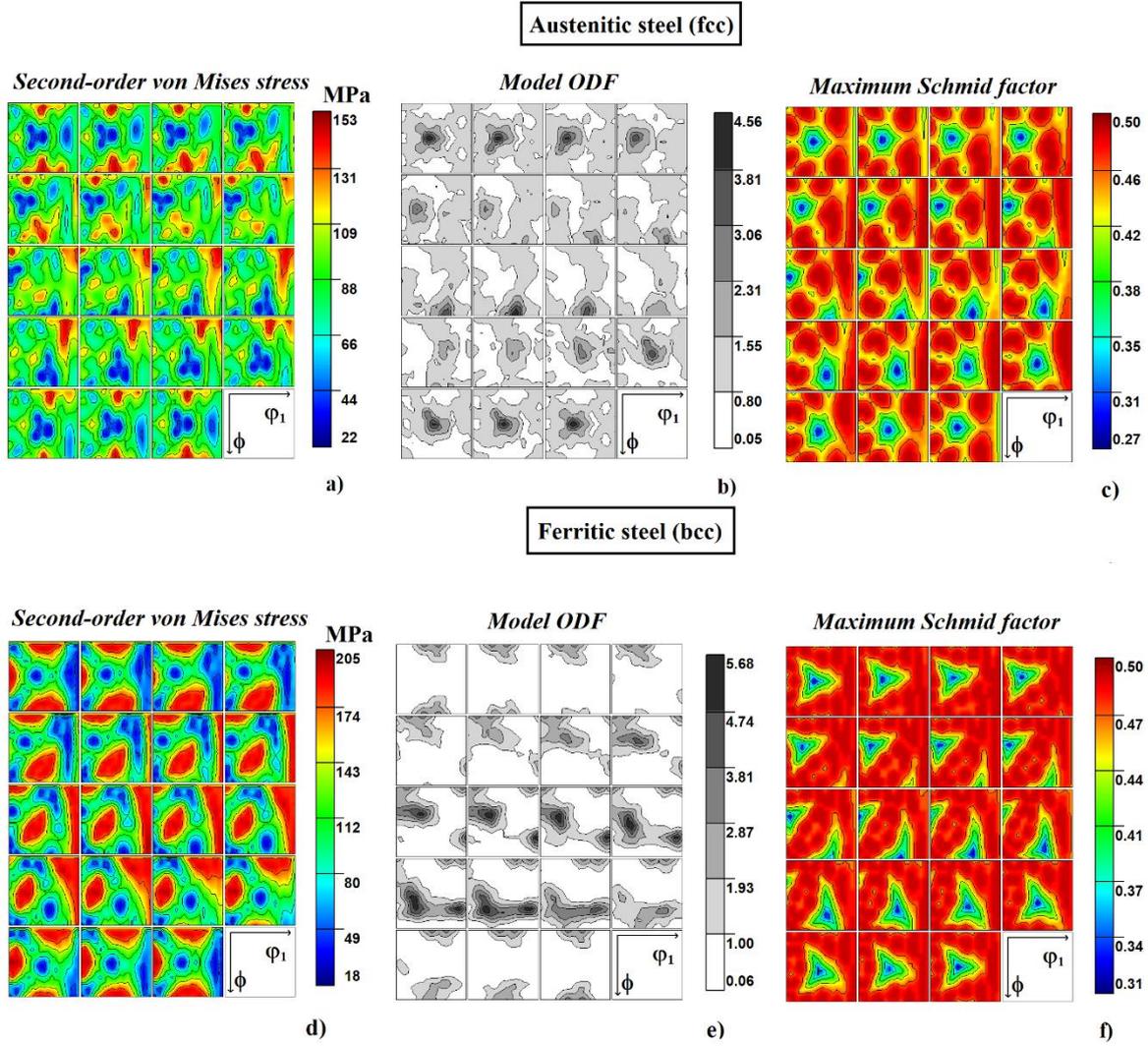

Fig. 15. Second-order plastic incompatibility stress (von Mises value $\sigma_{Mises}^{II,pl}$) obtained using the initial experimental texture and initial residual stresses (a, d), final ODFs obtained from model (b, e) and a maximum value of resolved shear stress (RSS) from all potentially active slip systems assuming uniaxial macrostress state (c, f). The results are presented in Euler space for austenitic (a, b, c) and ferritic (d, e, f) steels subjected to tensile deformation and unloaded.

## 4. Discussion

Based on the tests performed and the findings presented in section 3.1, it can be said that the Eshelby-Kröner model for the sample's interior, in which the lattice strains were measured for the elastic range of sample deformation, accurately predicts the XSF values (Fig. 6). This implies that the ellipsoidal Eshelby inclusion in the effective matrix representing the macroscopic sample can be used to predict the intergranular interactions, which are the elastic response of the grains to the applied load. For the tested textured samples, slight non-linearities of the determined $F_{11}$ vs. $cos^2\varphi$ plots were found when the load was applied along TD, and the orientation of the scattering vector changed between TD and RD. It was confirmed by both the experimental and calculated $F_{11}$ vs. $cos^2\varphi$ plots. An important takeaway of this result is that the non-linearities $<a(\psi,\varphi)>_{hkl}$ vs. $cos^2\varphi$ plots measured experimentally (Figs. 7-10) cannot be explained by the elastic anisotropy of the crystallites and textured sample.

After XSF verification, the methodology based on the results of the EPSC model was used to determine the first and second-order stresses in the initial and in situ deformed samples. The proposed approach of data interpretation made it possible to identify the reason of non-linearities in the $<a(\psi,\varphi)>_{hkl}$ vs. $cos^2\varphi$ plots, which can be explained as the effect of second-order plastic incompatibility stresses. Such stresses arise as a result of plastic anisotropy of individual grains, leading to mismatch with the neighboring ones and they can be simulated by the EPSC model. The very good agreement of the calculated and experimental non-linearities confirms the correctness of the model results.

Using the methodology described in the Introduction (Eqs. 5 and 6), a zero value of the plastic incompatibility was found for the initial austenitic sample and when the load was applied within the elastic range of deformation (two first points in Figs. 11 and 13 a). In this case, despite adjusting $q$, high values of $\chi^2$ were obtained, approximately equal to those obtained assuming $q = 0$ (cf. Fig. 13 a). However, this does not mean that the plastic incompatibility stresses $\sigma_{ij}^{II,pl}$ in the non-loaded sample are negligible, but their variation with orientation does not coincide with that predicted by the EPSC model. Therefore the stresses $\sigma_{ij}^{II,pl}$ cannot be determined for the initial sample and the sample subjected to the load within the elastic deformation range.

Then, when plastic deformation began, the value of $\chi^2$ decreased at about $E_{11}$ = 1.5 % (for the case of adjusted $q$) and stabilized at about $E_{11}$ = 1.5 % of sample strain. Starting from this point, the analysis of $\sigma_{ij}^{II,pl}$ can be performed because the fitted data matched the experimental points that was already observed for the example plots $<a(\psi,\varphi)>_{hkl}$ vs. $cos^2\varphi$ shown in Figs. 8 a and 10 a. Therefore, this means that the evolution of the value $\overline{\sigma_{Mises}^{II,pl}}$ was qualitatively determined, showing its progressive increase with plastic deformation (above approximately $E_{11}$ = 1 %) until the beginning of unloading, as presented in Fig. 11 b. Then the value of $\overline{\sigma_{Mises}^{II,pl}}$ stresses remains unchanged during sample unloading. It should be concluded that the sample strain of about $E_{11}$ = 1% - 2 % is enough to generate the incompatibility stresses $\sigma_{ij}^{II,pl}$ corresponding to tensile plastic deformation, which replaced the previous $\sigma_{ij}^{II,pl}$ stresses present in the initial austenitic sample. This result agrees well with the behavior of the second-order plastic incompatibility stresses determined recently during the tensile test in such materials as magnesium alloy [33], ferrite in pearlitic steel [32], and both phases in duplex steel [3].

Contrary to austenite, non-linearities of the $<a(\psi,\varphi)>_{hkl}$ vs. $cos^2\varphi$ plots measured in the initial non-deformed ferritic sample coincide well with model prediction causing significant improvement of the fitting, so the decrease of $\chi^2$, when the parameter $q$ is adjusted (Fig. 13 b). This significantly smaller value of $\chi^2$ for the analysis taking into account second-order stresses ($q$ adjusted) compared to the assumption of $q$ =0 is then observed during the elastic and plastic deformation of the sample. It was found that the character of the second-order stresses $\sigma_{ij}^{II,pl}$ in the non-loaded sample (subjected primarily to a cold rolling process) is similar to that which corresponds to tensile plastic deformation (as predicted by the model). It is also seen that the tensile test does not significantly change the character of the second-order stresses in the ferritic sample during the tensile test but the value of $\overline{\sigma_{Mises}^{II,pl}}$ increased during deformation (cf. Fig. 12 b).

Interesting results were obtained from the comparison of $\sigma_{Mises}^{II,pl}$ distribution with the ODF function (cf. Fig. 15). Despite the correlation between minima of $\sigma_{Mises}^{II,pl}$ with maxima of ODF in the case of austenite, such correlation was not confirmed by the ferritic sample. It leads to the conclusion drawn by Daymond et al. [42], based on lattice strain evolution in an austenitic sample during in situ tensile test, that the second-order stresses do not correlate strongly with the crystallographic texture. This fact can be explained by the Eshelby-type interaction in which the ellipsoidal inclusion interacts with the mean matrix representing polycrystalline material. Certainly, the deformation of the matrix depends on the texture, but this is not a very significant effect compared to differences in the plastic behavior of grains (inclusions), which are caused by the activation of slip systems depending on the lattice orientation with respect to the applied load. In turn, the resulting plastic incompatibilities between the grains and the matrix mostly depend on the RSS values on different slip systems. This was confirmed in the present work, where the correlation between the low values of the minimum Schmid factor (corresponding to uniaxial load) with the minima of $\sigma_{Mises}^{II,pl}$ stresses were found for both studied samples (cf. Fig. 15). Similar conclusions concerning the correlation between the Schmid factor for basal system and $\sigma_{Mises}^{II,pl}$ stresses was also observed for Mg-alloy studied in [18].

5. Summary and conclusions

This paper deals with the reasons for the non-linearity of the $<a(\psi,\varphi)>_{hkl}$ vs. $cos^2\varphi$ plots obtained by diffraction methods used for stress measurements. High energy synchrotron beam diffraction in transmission mode and prediction from the EPSC model allowed for studying the impact of crystallographic texture and second-order plastic incompatibility stresses on the results of stress measurements during in situ tensile tests carried out for ferritic and austenitic steels. The main findings of the work are the following:

- The $F_{11}$ XSF constants determined in situ from diffraction measurements carried out for the applied stress increments (elastic loading or unloading) are almost linear vs. $\cos^2\varphi$ when they are measured between TD and RD for both textured materials. This linear behavior was confirmed by models for XSF determination in which crystallographic texture is taken into account.

- The Eshelby-Kröner model used for XSF calculations agrees best with the experimental results.

- It is possible to determine the plastic incompatibility second-order stress only when the mode of deformation process is known. In this case, the model prediction of anisotropy of lattice strains due to second-order plastic incompatibility stresses is possible and the first-order as well as second-order stresses can be simultaneously determined.

- For the unloaded austenite sample, the non-linearities of the $<a(\psi,\varphi)>_{hkl}$ vs. $cos^2\varphi$ plots are small, while a significant undulation of the plots was observed for ferrite. For ferrite, it was possible to determine the second-order incompatibility stresses because the non-linearities correspond with the model prediction. For austenite, it was not possible because there are no such correlations.

- During the tensile test, the second-order incompatibility stresses remained unchanged in the purely elastic range, but they immediately transformed for the sample strain above 1 - 2%. Although these stresses changed significantly in the austenitic sample, such transformation did not occur for the ferritic sample. This is because the residual plastic incompatibility in the initial ferritic sample had a character similar to those generated during the tensile test.

- It was found that the second-order stresses are generated or modified only during plastic deformation, while they remain unchanged for a purely elastic sample deformation. The second-order incompatible stresses remaining in the samples after unloading as residual stresses were the reason for the non-linearities in the $< a(\psi, \varphi) >_{hkl}$ vs. $cos^2\varphi$ plots for both investigated samples.

- The orientation distribution of second-order plastic incompatibility stresses is not directly correlated with crystallographic texture but correlates with the maximum value of Schmid factor calculated for all potentially active slip systems.

**Acknowledgments**

We would like to thank Helmholtz-Zentrum Berlin für Materialien und Energie for the beamtime provided at 7T-MPW-EDDI beamline (BESSY II).

**Funding**

This work was financed by grants from the National Science Centre, Poland (NCN): UMO-2021/41/N/ST5/00394. Research project was partly supported by the program "Excellence initiative – research university" for the AGH University of Science and Technology.

**Conflict of interest**

The authors declare no conflict of interest.

**Authorship contribution statement**

M.M-W. Conceptualization, Methodology, Software, Validation, Data curation, Formal analysis, Investigation, Writing- Original draft preparation, Writing - Review & Editing, Project administration

A.B. Conceptualization, Methodology, Writing - Review & Editing, Funding acquisition

C.B. Sample preparation, Formal analysis, Review & Editing

M.W. Formal analysis, Investigation, Review & Editing

S.W. Investigation

R.W. Sample preparation

G.G. Investigation, Review & Editing

P.K. Software, Review & Editing

M.K. Conceptualization, Investigation, Methodology, Review & Editing

CH. G. Conceptualization, Investigation, Methodology, Review & Editing

**Data availability**

The raw data required to reproduce these findings are available to download from https://data.mendeley.com/datasets/trgrhm83hs/draft?a=a2b7f7eb-5533-4864-a5e8-22fc17babb0c. The processed data required to reproduce these findings are available to download from https://data.mendeley.com/datasets/b6v7w9rpdg/draft?a=f247ff02-764f-43e4-b59d-57ae4297fb49.